\documentclass{article}
\usepackage{graphicx} 
\usepackage{hyperref}
\hypersetup{colorlinks,allcolors=black}
\hypersetup{pdfborder=0 0 0}
\usepackage{amsmath}
\usepackage{amssymb}
\usepackage{geometry}
\usepackage{framed}
\usepackage[skip=10pt, indent=0pt]{parskip}
\usepackage{enumitem}
\usepackage{graphicx}
\usepackage[colorinlistoftodos]{todonotes}
\usepackage{graphicx}
\usepackage[T1]{fontenc}
\usepackage{tabularray}
\usepackage{xurl}
\usepackage{babel}

\begin{document}
\begin{titlepage}
\setcounter{page}{0}
\begin{center}
\large WITNESS REPORT

\vspace{5mm}

\LARGE
\textbf{TRIED: Truly Innovative and Effective AI Detection Benchmark, developed by WITNESS.}

\small
\textbf {{shirin anlen,}$^1$ {Zuzanna Wojciak}$^2$}\\[8pt]
 
\footnotesize
$^1$ Media Technologist at WITNESS, shirin@witness.org\\
$^2$ Program Consultant at WITNESS, zuzanna\_wojciak@witness.org\\[6pt]
\end{center}
\vspace{1cm}
\begin{abstract}
\small
The proliferation of generative AI and deceptive synthetic media threatens the global information ecosystem, especially across the Global Majority. This report from WITNESS highlights the limitations of current AI detection tools, which often underperform in real-world scenarios due to challenges related to explainability, fairness, accessibility, and contextual relevance. In response, WITNESS introduces the \textit{Truly Innovative and Effective AI Detection (TRIED) Benchmark}—a new framework for evaluating detection tools based on their real-world impact and capacity for innovation. Drawing on frontline experiences, deceptive AI cases, and global consultations, the report outlines how detection tools must evolve to become truly innovative and relevant by meeting diverse linguistic, cultural, and technological contexts. It offers practical guidance for developers, policy actors, and standards bodies to design accountable, transparent, and user-centered detection solutions, and incorporate sociotechnical conside-\\rations into future AI standards, procedures and evaluation frameworks. By adopting the TRIED Benchmark, stakeholders can drive innovation, safeguard public trust, strengthen AI literacy, and contribute to a more resilient global information credibility. \vskip 2mm


\end{abstract}
\vspace{55mm}
\begin{center}
\includegraphics[width = 40mm]{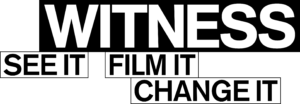}

\includegraphics[width = 50mm]{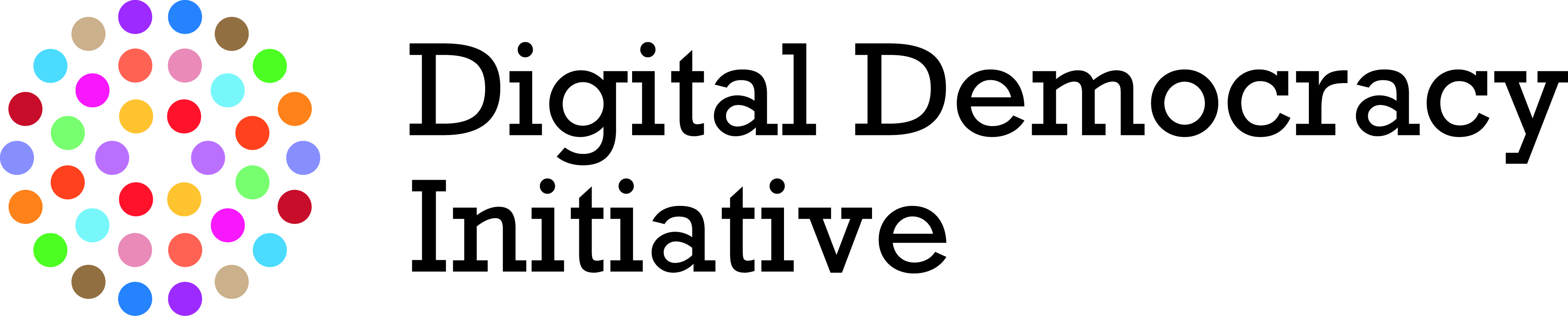}
\end{center}
\end{titlepage}

\newpage
\tableofcontents
\newpage
\section{Executive Summary}
The rapid rise of generative AI and the increasing ability to create deceptive synthetic media poses a significant threat to public trust and information integrity, particularly in resource-constrained regions of the Global Majority. WITNESS’ \cite{WITNESS}, \cite{TTO} work underscores the critical role of detection efforts, providing real-time crisis mitigation, enhancing media literacy, and strengthening public understanding of AI-manipulated content to protect trust in media. However, despite advancements in AI detection tools, their real-world effectiveness remains hindered by persistent challenges in accuracy, generalization, fairness, accessibility, transparency and explainability. These limitations prevent existing tools from being truly reliable in addressing real-world harms, especially in contexts where deceptive AI can have severe consequences.

To address the gap between the technical capabilities and practical applications of detection tools, this report introduces \textit{WITNESS’ Truly Innovative and Effective AI Detection (TRIED) Benchmark}—a framework that evaluates AI detection tools through a sociotechnical lens, focusing on adaptability, transparency, accessibility, contextual relevance, and fairness. By prioritizing these factors, the TRIED Benchmark ensures that detection tools are not only technically proficient but also innovative in how they address real-world needs, particularly in historically underserved regions and contexts. Key findings emphasize that for detection tools to be truly innovative and effective, they must analyze media in the formats and resolutions most commonly shared online, adapt to diverse linguistic and cultural contexts, invest in explainability, and integrate seamlessly into broader verification workflows. This approach advances detection technology beyond theoretical performance, making it impactful in the environments where it is needed most.

The report provides a detailed account of different dimensions of effectiveness supplemented by a list of practical considerations to guide innovative AI detection tools development as well as support policy efforts towards AI procedures and standards reflecting global frontline community input. Developers are urged to design adaptable, user-friendly tools that leverage diverse datasets, undergo regular updates, and provide clear, actionable insights. Standards bodies should establish guidelines on transparency, explainability, and system updates to align detection tools with sociotechnical evaluations and global best practices. International and domestic regulatory efforts should strengthen accountability structure, establish oversight mechanisms and governance, ensuring that future AI detection regulations reflect principles of trustworthy and human-centric AI and invest in media literacy to improve public understanding of AI-generated content. Governments and market leaders should increase investment in AI detection technologies and sociotechnical evaluations while reinforcing accountability frameworks, and support investment in training programs, workshops, and technical assistance to ensure stakeholders can effectively and responsibly leverage detection tools in real-world contexts––ultimately enabling the development of genuinely effective tools that provide real value to users.

By adopting the \textit{TRIED Benchmark}, stakeholders can create, innovate and assess detection tools that actually work in the real world that contribute to upholding trust, fostering accountability, and strengthening the global information ecosystem against the growing threats posed by deceptive AI.
\section{Introduction}
Introduction
Deceptive AI content continues to present a serious threat to society by undermining information integrity and public trust in the media, governments, and the work of civil society. Addressing the risks associated with generative AI requires detection solutions that not only excel technically but also perform effectively in diverse, real-world contexts. Detection tools for synthetic media can be highly effective when applied to well-structured, high-quality content. For example, WITNESS analyzed a deepfake video featuring a former advisor to the President of Ukraine \cite{facebook}. The video depicted the individual speaking directly to the camera, with clear audio and no background noise—conditions under which AI detection tools deliver highly accurate results. However, such scenarios are the exception rather than the rule. Most manipulated media online exists in far more complex and variable contexts, where challenges like poor-quality footage, layered manipulation, and linguistic diversity reduce the tools' effectiveness. Despite these challenges, the effectiveness of AI detection tools is often evaluated using technical metrics such as accuracy, speed, scalability, and generalization. While important, these metrics fail to address the complexities of applying detection tools in messy, real-world scenarios.

At WITNESS, initiatives like the Deepfakes Rapid Response Force (DRRF) \cite{DRRF}––which provides frontline information actors with access to AI detection tools and expertise––along with global consultations \cite{Fortifying}, have revealed a persistent “detection equity gap"\cite{Pre-Empting}. This gap highlights the disparity between the technical capabilities of AI detection tools and their practical utility in the contexts where they are most needed. The gap is particularly pronounced in the Global Majority, where resource constraints, issues of representation in tool development, linguistic diversity, and contextual manipulation trends exacerbate challenges. 
To address these issues, WITNESS proposes the \textit{Truly Innovative and Effective Detection (TRIED) Benchmark}. Informed by the needs and experiences of global actors most affected by deceptive AI and the weaponization of plausible deniability of authentic content, this report offers a comprehensive analysis of different dimensions of effectiveness of AI detection tools and outlines practical considerations to support the innovation of tools and the development of polices reflecting the input of the global frontline community. The report is supplemented by the \textit{Benchmark}, a detailed checklist designed to guide AI detection developers, and other stakeholders, in evaluating whether the detection tools align with best practices for effective detection.
While promising advancements in provenance standards and metadata-based verification exist, systemic issues in implementation and emerging challenges \cite{provenanceblog} of trust and digital divides leave universal feasibility of standards impossible. This means AI detection tools remain indispensable for verifying content integrity, exposing manipulation, and restoring trust in authentic media. These tools play a critical role in a comprehensive strategy to mitigate the risks of deceptive AI.

The \textit{TRIED Benchmark} promotes a resilient and public interest approach to assessing AI detection tools, going beyond technical specifications to address accessibility, transparency and explainability, fairness, contextual applicability, and the role of AI detection within a verification landscape. By prioritizing these dimensions, the framework equips developers and policy actors with criteria to design innovative tools that effectively tackle synthetic media in diverse, real-world environments and drive responsible, human-centric AI innovation \cite{innovation}. Only by addressing all aspects of effectiveness can AI detection tools fulfill their purpose: empowering communities worldwide to combat the growing threat of deceptive AI and safeguard information integrity.

To access the \textit{\textbf{WITNESS' TRIED Benchmark: A Checklist for Truly Innovative and Effective AI Detection}}, head to Annex A. 

\section{Background}
Since 2018, WITNESS has led a ‘Prepare, Don’t Panic’ \cite{preparedontpanic} approach to synthetic media, deepfakes, and multimodal generative AI. In response to the lack of access to reliable and transparent detection tools identified through this work, in March 2023, WITNESS launched the Deepfakes Rapid Response Force (DRRF), a pioneering initiative that connects frontline fact-checkers, journalists and civil society actors with leading media forensics and deepfake detection experts. This collaboration provides a timely, evidence-based analysis of content threatening democracy and human rights. 

The DRRF’s work is a testimony to the tangible impact of detection efforts. Its insights have helped mitigate crises in real-time, foster media literacy, and enhance public understanding of synthetic media. Cases analyzed by the DRRF have been widely reported, including by Human Rights Watch \cite{HRW}, the Rest of the World \cite{restoftheworld} and GHOne TV News \cite{GHOneNews}. Additionally, WITNESS has trained frontline journalists in AI detection and response strategies. These trainings reveal critical challenges faced by partners in the Global Majority, including fragile media ecosystems, ineffective detection tools due to gaps in training data for local languages, accents, and public figures, and the prevalence of manipulation tactics. Compounding these issues is a widespread lack of AI media literacy, which hinders the ability to interpret and trust detection outputs, and the trend to discredit authentic media content by claiming it had been AI-generated \cite{ISD}, further undermining the trust in the information ecosystem.

Strictly technical evaluations fail to capture these sociotechnical considerations, leading to an incomplete assessment of AI detection tools. A sociotechnical perspective expands the understanding of metrics of effectiveness by emphasizing the practical applications of AI detection tools in diverse contexts. They cannot be evaluated in a vacuum––they must be contextualized within these real-world applications and grounded in the lived experiences of those navigating these challenges. WITNESS' work highlights the importance of grounding assessments in the lived experiences of those navigating high-stakes scenarios. To provide adequate support to its users, AI detection approaches must address the needs and technical capacity of relevant actors \cite{Pre-Empting}, such as fact-checkers, frontline journalists, and civil society more broadly. A tool’s success should be measured not only by its technical performance but also by how well it aligns with the practical experiences and challenges of actors dealing with deceptive AI.

\section{Related Work}
As AI advances, its societal impact has come under increasing scrutiny. A growing body of work evaluates Al through a sociotechnical lens, exploring interactions between technology and society. This notion has been discussed in the context of safety \cite{sociotechsafety}, risks posed by synthetic content \cite{NIST}, algorithmic fairness \cite{unfairness}, and algorithmic detectability of generative AI \cite{governingaccess}. In policy, sociotechnical considerations have shaped frameworks emphasizing trustworthiness. For example, the European Union lists seven key requirements of a trustworthy AI system \cite{EU}, including transparency, robustness and fairness. Similar commitments are reflected by the National Institute of Standards and Technology (NIST) \cite{NISTtrustworthiness} and the Organisation for Economic Co-operation and Development (OECD) \cite{OECDAI} – a sentiment most recently embodied by Article 1 of the EU AI Act calling for trustworthy and human-centric AI \cite{AIAct}. 

These values extend to synthetic media and detection policies. Initiatives like the Partnership on AI (PAI) outlined essential insights and recommendations for the field following the Deepfakes Detection Challenge in 2020 \cite{PAIchallenge}, as well as highlighted the need for a responsible and transparent approach to detection \cite{PAIchallenge} and provenance \cite{PAIglossary} and together with WITNESS, advocated for equitable access to AI detection solutions \cite{governingaccess}. There has been considerable progress in transparency methods as a way to ensure the authenticity of media, through initiatives such as the Coalition for Content Provenance and Authenticity (C2PA) \cite{C2PA} and the incorporation of a sociotechnical lens in the C2PA's Harms Modelling analysis \cite{C2PAharms}. 

With regard to detection, prevailing approaches to evaluation of the AI detection tools often do not account for their performance on real-world examples of deepfakes. Deepfake-Eval-2024 Benchmark \cite{deepfakeeval2024} and the work by the Commonwealth Scientific and Industrial Research Organisation (CSIRO) and South Korea's Sungkyunkwan University \cite{Sok} are recent efforts that highlight the shortcomings of AI detection models when applied in real-life scenarios and call for a more comprehensive approach to assessments considering different factors that may limit the effectiveness of the tools. However, while an important step towards more effective AI detection tools, these works are still primarily technical and do not include broader sociotechnical considerations, such as accessibility, fairness and explainability.  

To address this existing research gap and embed sociotechnical perspective in the evaluations of AI detection tools, we've developed the \textit{TRIED Benchmark}, grounded in established principles of trustworthy AI and informed by our work with the DRRF. This approach goes beyond abstract notions of trust, offering a practical methodology to evaluate AI detection tools against their effectiveness and impact in diverse contexts. By providing concrete guidance, we aim to bridge the gap between ethical AI commitments and real-world implementation, empowering policy actors and AI developers to ensure responsible and trustworthy deployment of these technologies.

\section{Redefining Innovative Effectiveness of AI Detection Tools in Practice}
Incorporating sociotechnical considerations into the evaluation of AI detection tools requires asking critical questions about the objectives and characteristics of a reliable AI detection system. The questions should explore:
\begin{itemize}[noitemsep]
    \item \textbf{Desired outcomes}: for example, what specific goals is the tool designed to achieve? How well does it perform across diverse real-world scenarios? Is its effectiveness aligned with the needs and experiences of the intended audience?
    \item \textbf{Unintended consequences}: for example, what potential risks or harms might arise, particularly for marginalized communities? How effectively are the tool’s limitations and intentions communicated to users?
    \item \textbf{Implementation guidelines}: for example, how do elements constituting trustworthy AI translate into practical considerations? How can the developers incorporate notions such as fairness, transparency, robustness, and explainability into the development and deployment of the AI detection tools? How to ensure that the end users understand and trust the outputs provided? How effectively does the tool integrate into existing skills and verification workflows? Does the tool complement the verification process without positioning itself as a definitive solution? Are results communicated in ways that add value to broader investigations?
    \item \textbf{Diversity in development}: for example, is the tool trained and tested on diverse datasets that reflect different contexts, characteristics, and media types? Does its performance remain balanced across diverse groups, or are its limitations in diversity clearly communicated? Does the evaluation process include diverse perspectives in testing and validation to ensure its effectiveness in real-world applications?
    \item \textbf{Distribution of benefits}: for example, are the tools equally useful to different audiences? Are they accessible and affordable to diverse users, considering barriers such as languages, diverse levels of technical skills, and access to computational knowledge and resources?
    \item \textbf{Proactive approach}: for example, what mechanisms are in place to identify and address the tool’s limitations or failings revealed in the course of the tool’s application with in-the-wild content? Is the tool designed with long-term effectiveness in mind? How is the tool future-proofed for continuous improvement of audiovisual AI and the development of other detection tools? How is transparency balanced with protecting proprietary technology, information security, or preventing bad actors from reverse-engineering the tool?
    \item \textbf{Accountability and governance}: who is responsible for oversight and accountability in the tool’s deployment, implementation, future updating, and retirement process? How are ethical concerns addressed during the tool's lifecycle, from development to deployment, updating and retirement process? Are mechanisms in place to address misuse or abuse of the tool?
\end{itemize}

\textbf{Based on these considerations, we propose six core pillars for evaluating innovative effectiveness in AI detection tools.} These core considerations highly influence each other and are interconnected. A comprehensive approach to sociotechnical evaluation of AI detection tools has to account for balancing tradeoffs among these considerations.

\subsection{Design AI Detection Tools to Deal with Challenges Posed by Synthetic Media in Real-World Scenarios.}

Key considerations for designing AI detection tools to deal with real-world scenarios: 
\begin{itemize}[noitemsep]
    \item \textbf{Adaptability to novel cases}: How well does the tool handle emerging and unexpected types of synthetic media? Is the team equipped to continuously update and maintain the system in response to evolving data and shifting outputs? Do they have a structured framework for conducting periodic evaluations to assess performance, accuracy, and the impact of false positives and false negatives over time?
    \item \textbf{Pre-release testing}: Was the tool tested on diverse “in-the-wild” content before deployment?
    \item \textbf{Performance on low-quality media}: How well does the tool process low-resolution, compressed, or heavily formatted files?
    \item \textbf{Handling diverse audio artifacts}: Can it reliably detect manipulation in recordings with background noise, music, or overlapping voices?
    \item \textbf{Diverse training data}: Was the tool trained on varied file types, compressions, and social media and messaging apps’ formats?
\end{itemize}
Effective AI detection tools must handle the complexities typical of audiovisual AI in real-world scenarios. Insights from the DRRF highlight these challenges. Tools tend to perform best on high-resolution and near-original materials, yet much of the content analyzed by fact-checkers, journalists and human rights actors contain dynamic environments, noisy backgrounds, and high compressions due to social media and messaging apps. These platforms may re-encode a file’s metadata at the time of uploading. Due to this encoding, social media platforms inadvertently create versions \cite{MagnetForensics} from user-submitted images and videos for widespread dissemination. These reprocessed files are often of lower quality and may include additional data introduced by the platform itself. This gets even more complicated when content is shared and moves quickly across platforms, further altering its properties. These changes significantly hinder AI media detection, making it one of the biggest challenges for experts analyzing digital content. Some DRRF experts did not engage with cases where the quality of content was too low to provide reliable results. For example, a suspected AI video from the Philippines purportedly showed the president snorting cocaine \cite{phillipines-cocaine}. The video’s poor quality rendered AI detection tools ineffective. It is key that such a common feature should not be overlooked and that the tool offers solutions to address this challenge. Since low-quality, low-resolution, and high-compression settings can obscure deepfake artifacts, AI detection tools must not only clearly communicate their limitations but also incorporate strategies to mitigate these issues. For instance, providing detailed explanations of these constraints and integrating support for complementary fact-checking and verification processes is essential for ensuring a comprehensive and trustworthy analysis (more on these strategies in the following sections). These shortcomings were also highlighted by the NIST – improving the performance of tools on post-processed content or media corrupted by noise or formatting \cite{NIST} can be seen as opportunities to further develop and improve the AI detection solutions to ensure they can provide adequate support to actors relying on them. 

Background noise, music, or cross-talking in audio recordings further complicate the detection process and lead to inconclusive results. For example, in an audio clip allegedly featuring the President of Russia \cite{Putintiktok}, the DRRF experts noted that the presence of only one speaker and the lack of background noise increased the reliability of the detection results. In contrast, loud background music in an audio recording allegedly featuring the voice of the President of Georgia \cite{georgiaprez} obscured crucial artifacts, complicating analysis. Similarly, in a case involving a purported phone call between an Indian politician and a YouTube influencer \cite{Indiaaudio}, overlapping voices hindered the detection process. The presence of multiple voices can hide signs of manipulation. Similarly, a recording including a mix of manipulated and authentic voices can also confuse the detector into misclassifying the recording as a whole as either manipulated or authentic. These scenarios illustrate the need for tools to perform reliably on noisy or complex audio recordings, or ideally offer mitigation strategies as mentioned above.

The effectiveness of detection tools depends in great part on whether training datasets represent the type of content they are expected to analyze. Researchers behind the Deepfake-Eval-2024 Benchmark \cite{deepfakeeval2024} observed how both open-source and commercial detection models performed poorly on real-world examples of deepfakes. However, the performance of open-source models significantly improved once they had been finetuned on a dataset representative of diverse in-the-wild deepfakes. The quality of detection results directly depends on whether the kind of content uploaded to the tool was present in the training dataset. For instance, to detect manipulation in low-resolution images, such images need to be included in the training data. Datasets should encompass diverse file types, augmentations, social media and messaging apps formats, and use cases to improve detection outcomes. 

The diversity of types of media submitted by the fact-checkers presents another challenge to the detection tools. For example, the DRRF received a suspected radio conversation recording (walkie-talkie) from the Sudanese civil war \cite{Sudanaudio}. Detection models failed to conclusively analyze the recording since radio conversations often appear at a different frequency than the data used for training, resulting in fundamentally different outcomes. As AI continues to evolve, forensic experts are likely to encounter novel forms of synthetic media. While no tool can fully anticipate future deceptive content, developers should prioritize adaptability, enabling tools to collaborate with users to handle unexpected cases. Some detection tools can adjust their performance based on user input. For example, a person-of-interest detection system analyzes uploaded content by comparing it with verified samples of a specific public figure \cite{Person-of-Interest}. Such a technique can adapt in real time to what person it will be analyzing and its accuracy may depend on the amount of data uploaded by the user. It demonstrates a possibility of creating detection tools adaptable to specific content and detection aims. However, it is equally important to communicate the tool’s limitations to the user if the tool was not trained on, and as a result is not intended to work, on a certain type of data (discussed more in the following sections). 

Another recurring challenge involves authentic content being falsely discredited as AI-generated. For instance, a video of Ghana’s Vice Presidential Candidate giving a public speech \cite{DadzieTV} was dismissed as a deepfake by the speaker himself. Although DRRF experts confirmed the video’s authenticity, providing evidence-based analysis to support this conclusion was challenging—it is inherently more difficult to prove lack of manipulation than to identify manipulation. These examples expose the gap between the public needs and current detection capabilities, underscoring the importance of designing tools that respond effectively to real-world scenarios. 

\subsection{Promote Transparency and Explainability of the Detection Results in a Clear and Detailed Manner.}

Key considerations for promoting transparency and explainability: 
\begin{itemize}[noitemsep]
    \item \textbf{Resource investments}:
    \begin{itemize}
        \item Are developers prioritizing resources to provide clear and evidence-based results?
        \item Are efforts directed at ensuring transparency and accessibility?
    \end{itemize}
    \item \textbf{Depth of information}:
    \begin{itemize}
        \item What information does the tool provide when sharing the outcome with the user?
        \item Does the tool offer detailed insights, such as evidence of AI manipulation?
        \item Does the tool offer insights on the methodology behind the scoring and the parameters that were set to generate results?
        \item Are limitations, such as dataset gaps or media quality issues, disclosed?
    \end{itemize}
    \item \textbf{Accessibility of communication}:
    \begin{itemize}
        \item Is the information presented in a clear and user-friendly manner?
        \item Are technical details simplified without losing critical nuances?
    \end{itemize}
    \item \textbf{Disclosure of limitations}:
    \begin{itemize}
        \item Are the tool’s limitations clearly communicated?
        \item Can users easily understand these limitations when interpreting results?
    \end{itemize}
    \item \textbf{Stated objectives}: Are the tool’s purpose and scope clearly defined?
    \item \textbf{Language used}: Does the tool avoid binary labels like “real” or “fake” or other terms that reflect false confidence and imply all AI-generated content is inherently deceptive, instead offering nuanced descriptions that reflect the complexity of the evidence?
    \item \textbf{Support materials}: Is the user provided with additional resources that would help them better understand and use the tool, such as a user guide and case examples of previously analysed content?
\end{itemize}
The effectiveness of an AI detection tool heavily depends on the clarity, details, and quality of the information it provides. Transparent and detailed communication empowers users to responsibly interpret detection results and fosters trust in the tool. This requires a user-centric approach, grounded in humility and an awareness of the tool’s capabilities, limitations, and intended audience.

Detection tools should explicitly define their strengths—such as the types of content they can analyze—and their limitations, including language constraints or contextual challenges. If tasked with analyzing content outside their training scope, tools should clearly communicate that the results may be unreliable. Explainability of the tool rests on the annotation of the training data set, which is a time-consuming and resource-intensive process, in particular in the context of unexpected cases as described above. Therefore, it is important to be mindful of the effort required to ensure transparent results. 

Many tools present detection results as confidence scores or binary outputs \cite{reutersarticle}, which, without additional context, can be difficult to interpret and may even create further confusion. For example, fact-checkers frequently report difficulty interpreting confidence scores, limiting their utility in verification processes. To improve usability, tools should provide detailed explanations of their processes, including:
\begin{itemize}[noitemsep]
    \item \textbf{Training data and objectives}: Top level information of the types of datasets were used and what types of manipulation the tool is optimized to detect.
    \item Limitations: How factors like low-quality media or background noise may affect outcomes.
    \item \textbf{Temporal localization}: Identifying specific moments and regions where manipulation occurred.
    \item \textbf{Provenance information for analyzed content}: Metadata insights such as timestamps, URLs, or accounts linked to the analyzed content (while ensuring the privacy of content creators).
    \item \textbf{Tool updates}: Information about the tool’s latest updates, reflecting ongoing improvements (importance of which will be discussed in more depth in the durability section).
\end{itemize}
NIST’s ‘Reducing Risks Posed by Synthetic Content’ report underscores the importance of supplementing detection results with relevant contextual information \cite{NIST}. Tools should highlight specific manipulated areas within content or provide metadata-based insights. It is key to note that the information provided to support the detection result should not include information on the actors who created and shared the content to protect their right to privacy \cite{ticks}. 

There is a recognized tension between transparency and security. Providing granular details on certain aspects of these models, such as the exact training datasets or detection methodologies, may expose them to adversarial attacks. Research indicates that detailed knowledge of a model's training data can enable adversaries to craft targeted inputs that exploit vulnerabilities, bypassing detection mechanisms, rendering detection tools less effective over time \cite{governingaccess}. This concern has been widely discussed in security research \cite{AdversarialMachineLearning, thesisRigaki, AdversarialMLAttacks}, particularly in adversarial machine learning, where attackers refine synthetic media to bypass known detection mechanisms. To mitigate this risk, we recommend sharing high-level information about detection models, such as their design objectives or the types of media they are trained to detect, rather than specific manipulation patterns or source code. It is vital that governments and regulators understand this inherent tension. Overly stringent requirements for granular transparency, without considering the security implications, could inadvertently weaken the very tools designed to combat misinformation. A nuanced approach that prioritizes high-level transparency and user understanding, while safeguarding against adversarial exploitation, is essential for the responsible deployment of AI detection technologies.

The need to provide the user with a comprehensive information on the tool's model level has been reflected in model cards \cite{modelcards} and datasheets for datasets \cite{atasheetsdatasets}. Such records detail the background information on the AI model and its workings, which allows different stakeholders to better understand and develop applications. Similar factsheets could accompany the detection tools adding application level and outlining additional information, including the tool's intended use, limitations, and evaluations across different factors. Currently, no settled guidelines exist for what detection tools should communicate, creating inconsistencies. There is an urgent need for establishing standards, such as uniform adoption of factsheets based on the framework of model cards and datasheets, which would ensure explainability, prioritize transparency, and improve user trust. Detection tools must also adapt language that reflects the complexities of AI-generated content \cite{leibowicz2025regulatingrealityexploringsynthetic} and moves beyond simplistic binary labels like "real" or "fake". This nuanced approach is essential in today’s rapidly evolving media landscape, where AI-generated content doesn’t inherently equate to "fake" and where clear, detailed communication is vital for fostering trust and usability.

WITNESS’ DRRF exemplifies this approach by tailoring technical insights from experts into accessible and transparent language for fact-checkers. These explanations emphasize processes and limitations that may affect results, particularly in challenging cases where multiple tools produce conflicting outcomes. For example, in cases from Nigeria \cite{Bigeria-twitter}, Uganda \cite{Kizza}, and Venezuela \cite{venezuela}, low-quality recordings led to inconclusive results, undermining the tools' reliability. Addressing such challenges requires transparency about how specific factors, such as poor quality or background noise, may have influenced outcomes.

Providing users with clear explanations and transparent breakdowns of the detection process not only builds trust but also enables information intermediaries to educate broader audiences about generative AI and deceptive media. This fosters responsible communication, bridges gaps in AI media literacy—especially in underrepresented regions—and supports informed public discussions.
\subsection{Ensure that AI Detection Tools are Accessible to Their Target Users.}
Key considerations for ensuring accessibility: 
\begin{itemize}[noitemsep]
    \item \textbf{Technical knowledge}: Can users with basic technical skills confidently use the tool and interpret results?
    \item \textbf{Affordability}: Is the tool free to use, or is it priced affordably for its target audience?
    \item \textbf{Offline functionality}: Can the tool operate in areas without a stable internet connection?
    \item \textbf{Data privacy}: Does the tool communicate clearly its data retention, storage and disposal policies to ensure that data privacy concerns do not deter users from uploading content containing personal or potentially sensitive information?
\end{itemize}
An effective AI detection tool should be accessible to its intended users \cite{whats-needed-in-deepfake-detection}, including communities with limited resources. While a tool may function in theory, barriers such as language limitations, technical skill requirements, connectivity issues, data privacy concerns, and costs can hinder its usability. Developers must clearly define their target audience and ensure the tool meets their needs.

If the tool is to be used by a wide audience, it should accommodate users with varying levels of technical expertise. For instance, a voice recognition tool requiring Python programming knowledge was deemed exclusionary by a fact-checker. Additionally, and in line with the explainability considerations above, detection outcomes must be communicated in a clear and straightforward language to make the tools accessible to a non-technical audience.

Interoperability is another crucial factor. Some detection tools only analyze content generated with specific technologies, forcing users to navigate a fragmented landscape of AI techniques. This not only limits effectiveness but also adds unnecessary burdens on users. 

Connectivity and privacy challenges further impact accessibility. Tools reliant on stable internet connections or those with high storage and processing requirements may exclude users in areas with unreliable connectivity. Offline functionality would significantly enhance usability in these contexts. Data privacy and security concerns also affect accessibility. Lack of clear and transparent policies on data retention, storage and disposal procedures may deter users from uploading content containing personal or potentially sensitive information.

Cost remains a significant barrier. Paywalls—whether per verification or via subscriptions—restrict access, especially for resource-constrained communities. Notably, paid tools do not always guarantee reliable results. Researchers also noted a trend where free or low-cost tools become more expensive once the service acquires a big user base. Such practice again limits the tool's accessibility and has particularly detrimental effects if the journalists, fact-checkers, and human rights actors had consistently relied on the tool in their workflow.   

However, accessibility without a robust implementation can amplify the risks posed by deceptive AI. For example, cases when online deepfake detectors incorrectly flag authentic media as fake, such as a video of Phyo Min Thein \cite{the-world-needs-deepfake-experts}, the former chief minister of Yangon, Myanmar. DRRF also dealt with a case from Mexico \cite{chapo} where journalists tested two images—an original photo and a deepfake—using several publicly available detectors. The tools produced conflicting results, with one even identifying the original photo as manipulated. In comparison, DRRF’s teams have access to state-of-the-art tools and obtained a conclusive result that the image was manipulated. Publicly available deepfake detectors can also easily be tricked into providing incorrect results \cite{reutersarticle}. For example, cropping or lowering the resolution of an AI-generated image caused a detector to falsely identify it as authentic after initially labeling it as fake. Similarly, screenshots of AI-generated images or recordings of AI-generated audio often confuse these tools, as such transformations strip critical information used for verification. 

These examples highlight the urgent need for advanced, reliable detection tools tailored to the specific contexts and requirements of fact-checkers, journalists, human rights defenders, and civil society more broadly. Such tools must be designed to handle real-world challenges effectively, fostering trust and ensuring their utility across diverse use cases.

\subsection{Embed Fairness Throughout the Process of Development and Distribution.}
Key considerations for embedding fairness:
\begin{itemize}[noitemsep]
    \item \textbf{Diversity in training data}: Is the dataset representative of diverse demographics, languages, and contexts?
    \item \textbf{Ethical collection of training data}:
    \begin{itemize}
        \item Were the datasets collected responsibly and ethically?
        \item How does the source of funding influence data collection practices and fairness?
    \end{itemize}
    \item \textbf{Compliance with relevant data privacy regulations}: Does the data collection process adhere to the local data protection, intellectual property, and biometric data laws?
    \item \textbf{Holistic testing approach}:
    \begin{itemize}
        \item Are external, diverse teams involved in testing?
        \item Does testing account for regional, cultural, and linguistic contexts?
    \end{itemize}
    \item \textbf{Performance equity}:
    \begin{itemize}
        \item Does the tool perform consistently across different demographics and contexts?
        \item What measures are in place to address disparities in performance?
    \end{itemize}
    \item \textbf{Contextual generalizability}:
    \begin{itemize}
        \item Can the tool reliably function across diverse regions, languages, and cultural contexts?
        \item Are steps taken to include content from underrepresented communities and contexts?
    \end{itemize}
    \item \textbf{Proactive identification of blind spots}: Are there established processes to detect and address potential blind spots effectively?
    \item \textbf{Team diversity}:
    \begin{itemize}
        \item Does the development team reflect the diversity of the communities the tool aims to serve?
        \item Are regional, cultural, and linguistic perspectives incorporated into the design process?
    \end{itemize}
\end{itemize}
Fairness in AI detection tools must be prioritized at every stage—from development to deployment—and encompass the distribution of benefits to diverse communities. Fairness in training data is foundational, as it directly impacts detection accuracy and equity. Research analyzing fairness in deepfake detection indicates how the distribution of attributes, such as race or gender, in the training dataset affects the performance of the AI detection tools \cite{deepfake-detection-improves}. 

For example, DRRF encountered a video of a Ghanaian politician \cite{ghana} where detection was hindered due to darker skin tones and high video compression. Having analyzed the datasets of popular deepfake detectors, experts also noted how the datasets largely consisted of Caucasian faces \cite{ju2023improvingfairnessdeepfakedetection}. This led to disparities where content featuring other demographics was more likely to be labeled as fake. Lessons from early implementations of facial recognition technologies (FRT) underscore these risks \cite{face-recognition-nist}. Unbalanced training datasets led to FRT disproportionately misidentifying people of color, exposing them to discrimination and harmful consequences \cite{AJL_facial_recognition}. Similarly, biased AI detection tools can produce unreliable results, amplifying human rights risks and disproportionately affecting already vulnerable communities. However, it is key to note the surveillance and human rights-related risks  stemming from creating extensive image databases \cite{Murray}, in particular considering the more and more prevalent use of facial recognition by the police \cite{big-brother-facial-recognition} and their growing access to databases not designed for policing purposes \cite{big-brother-watch-clause21}. To address these concerns, both fairness and privacy must be central considerations throughout the development of detection tools.

Another challenge is the lack of data representing linguistic diversity and local contexts. Many tools are trained on limited data, such as English and Spanish-language content or "talking head" videos, such as those found in datasets like FaceForensics++ \cite{FaceForensics++}. This narrow focus limits their effectiveness in analyzing diverse languages, accents, and content types. For instance, in a Sudanese case \cite{sudan-Abdel-Fattah-al-Burhan}, DRRF experts struggled to verify a conversation recording due to the tool’s inability to process the local language. This creates a significant barrier for communities from the Global Majority operating in underrepresented languages, who may be reluctant to use them. Similarly, identity-based tools requiring reference samples are ineffective when public content of the individual is scarce or does not exist at all, as seen in a case involving a local journalist from Venezuela \cite{venezuela-audio}.

The data used for training should be sourced in a fair and responsible way, including with the consideration of intellectual property, data protection and biometric data laws. When creating and using training datasets, developers should ensure compliance with local data protection regulations, particularly with regards to the lawful grounds for data processing and safeguards against data leaks of sensitive information. It is worth considering how the funding of the tool can affect the type of data used for the training. 

Ensuring fairness also requires rigorous and inclusive testing processes. Developers should involve external teams from varied cultural, regional, and linguistic backgrounds to identify vulnerabilities and address blind spots. Global red-teaming efforts can ensure tools perform reliably across different contexts. Marketing of the tool also requires a fair and humble approach – the tool’s accuracy should not be exaggerated \cite{FTC} and any claims should be supported by explanation so that information actors can rely on them in a responsible manner.  

\subsection{Invest in Detection Durability.}
Key considerations for investing in detection durability:
\begin{itemize}[noitemsep]
    \item \textbf{Regular testing and updates}:
    \begin{itemize}
        \item Is the tool tested regularly and updated frequently?
        \item Does funding impact the frequency or quality of updates?
        \item What support systems ensure consistent maintenance?
    \end{itemize}
    \item \textbf{User communication:}
    \begin{itemize}
        \item Are users informed about updates and their impact?
        \item Is the tool’s current status, including limitations, clearly communicated?
    \end{itemize}
   \textbf{ \item Adaptability to new technologies}:
    \begin{itemize}
        \item Can the tool easily adapt to and incorporate new technology advancements?
        \item How flexible is the tool in addressing evolving forms of AI?
        \item Are measures in place to maintain performance across diverse AI-generated or edited content?
    \end{itemize}
    \item \textbf{Responsible maintenance and defined retirement protocol}:
    \begin{itemize}
        \item If updates cease, is the tool responsibly removed from public access?
        \item Are processes in place to prevent outdated tools from producing false or misleading results?
    \end{itemize}
    \item \textbf{Future proofing}:
    \begin{itemize}
        \item How are risks from adversarial attacks continuously assessed and mitigated after a tool's release?
        \item How do these efforts build on safeguards established during development and risk assessments conducted prior to launch, ensuring a consistent and proactive approach to long-term resilience?
    \end{itemize}
\end{itemize}

As synthetic media grows increasingly sophisticated, AI detection tools must be regularly updated, evaluated and maintained to remain effective. Durability in detection tools is not just about maintaining functionality but ensuring their relevance in an ever-evolving landscape. This requires routine testing, adaptability to emerging technologies, and resilience against adversarial attacks. Consistent performance evaluations, such as reassessing previously verified cases and tracking accuracy over time, as well as continuous review of new classification tools and techniques, are critical. Developers should also anticipate advancements in AI and design flexible systems capable of integrating new techniques.

The creation and maintenance of AI detection tools should also contain an element of anticipation\\—understanding the generative AI field and the pace at which technologies advance should prompt them to consider how the field and tech will change in the following years. Developers should future-proof their tools to guarantee that their approaches remain relevant in the long term and provide reliable results to those who depend on them. This includes preparing for long-term exposure to adversarial attacks, which can undermine the tools’ reliability and model’s security. These efforts require substantial funding. Limited financial resources can restrict the frequency and scope of updates, impacting the long-term effectiveness of detection tools. Alternative funding models and regulatory support are essential to ensure that tools are maintained responsibly and provide reliable performance to those who depend on them.

It is equally important that the current status of the AI detection tool is communicated to the user - a tool that has not been updated for a certain period may fail to detect newer cases of synthetic media or provide inaccurate results. In the DRRF, experts inform the WITNESS team about updates and their impact on tool capabilities. However, some fact-checkers admitted they overlook the role of maintenance when evaluating tools. If the tool does not seem useful, they will exclude it from their verification process and will not return to use it. To prevent this, developers must actively notify users about updates and improvements. Tools should have a well-defined retirement protocol in place before their release. If a tool is no longer maintained, it should be removed from public access \cite{mm-maybe-AI-image-detector} to prevent the risk of generating false outcomes. Ensuring transparency about the tool’s status and updates fosters trust and encourages continued use by fact-checkers and other stakeholders.

The introduction of durability-related standards should address some of these challenges. Clearly established timelines for maintenance and criteria for removing outdated tools can help ensure reliability. Collaborative benchmarks comparing different detection solutions can also support the development of resilient systems capable of adapting to advancements in generative AI.   

\subsection{Leverage and Support the Integration of Tools Within a Broader Verification Ecosystem.}
Key considerations for leveraging and supporting integrating detection tools within a broader verification ecosystem:
\begin{itemize}[noitemsep]
    \item \textbf{Integration with the broader ecosystem}:
    \begin{itemize}
        \item How well does the AI detection tool fit within a broader information verification workflow?
        \item Can it be easily combined with other journalistic or verification methods?
    \end{itemize}
    \item \textbf{Supportive role}:
    \begin{itemize}
        \item Does the tool complement the verification process without positioning itself as a definitive solution?
        \item Are results communicated in ways that add value to broader investigations?
    \end{itemize}
    \item \textbf{The role of context}:
    \begin{itemize}
        \item Are tools evaluated based on their performance in human-assisted detection workflows?
        \item How is the importance of context integrated into the detection process?
    \end{itemize}
    \item \textbf{Primary use}: What additional support and information could the AI detection tool provide in scenarios when it is the primary solution?
\end{itemize}
A robust evaluation of AI detection tools must position them as part of a broader verification landscape, examining their complementary relationship with other methods. AI detection tools alone do not offer a definitive answer to confirm or deny the authenticity – rather they can provide a piece to a larger verification puzzle which consists of other crucial techniques such as open-source methods and tracking relevant contextual information.

Use of other verification methods alongside the detection tools is particularly important in cases where the latter cannot provide confident results. Considering that they rarely work with high-resolution original content, journalists and human rights researchers often resort to other techniques, such as using satellite imagery analysis, chronolocation, witness testimonies or looking for connected footage to verify a piece of media.  WITNESS’ Guide to Community-based Approaches to Verification \cite{community-based-approach-to-verification} lists diverse verification processes and tools to support individuals in asserting credibility of visual media. Fact-checkers also emphasize that AI detection tools are just one step in their verification workflows, often supplemented by additional tools and approaches. 

Similarly, in analyzed cases, DRRF teams have often recommended using other methods to corroborate deepfake detectors’ results. For example, in one specific case \cite{Ukrainian-soldier} where various AI detection tools provided conflicting results, the teams resorted to manual verification methods such as tracking related accounts, identifying similar image styles, and leveraging additional image analysis tools to analyze the image in question and were able to assert that it was authentic. The availability of diverse tools is particularly important in cases where the information actors aim to prove authenticity of a piece of media that is falsely purported to be AI-generated. As discussed in the previous sections, it is far more challenging to prove absence of manipulation than its presence. In such cases, the ability to support AI detection analysis with other tools and techniques provides users with more ways to assert credibility of a piece of media.

The need for reliance on diverse verification approaches is further underscored by the public confusion around the capabilities of AI detection methods. Due to lack of general understanding of synthetic media and AI detection techniques, some users approach the tools believing that they can detect any type of manipulation, including whether a piece of media was manually edited. Additionally, the evidence of AI manipulation itself is sometimes seen as evidence of intention to deceive, even when it does not affect the content of the message. For example, the use of AI-generated voiceover in a video does not automatically determine that the content of the message is also inauthentic; instead, the user should refer to other sources and methods to verify it. In a similar manner, researchers at Google highlight the need for a dual approach to verification \cite{Google-Public-Policy}, including both an analysis of how a piece of media was created or edited, and examining the background information and the context in which a piece of content was shared to fully evaluate its trustworthiness.  

Given the diverse needs of users, specialized tools tailored to specific audiences, such as human rights defenders and journalists, can significantly enhance verification efforts. DeFake.app \cite{de-fake} is an example of a detection tool created through a collaboration with journalists, and tailored to their specific needs. Other tools, such as Amnesty International’s Youtube Data Viewer \cite{youtube-data-viewer}, designed to extract metadata from YouTube videos, further demonstrate the value of tools created with specific use cases in mind. Similarly, collaborative initiatives like the Shakti Collective \cite{Shakti} in India offer valuable insights into fact-checking workflows in complex, multilanguage information ecosystems, providing guidance for developing tools that address diverse use cases.  

AI detection tools play a vital role in helping fact-checkers verify and communicate the authenticity of content to the public. Tools that provide clear, evidence-based analysis strengthen the credibility of the verification process and enable fact-checkers to confidently and safely share their findings. For example, DRRF analyses have supported reporting on alleged deepfakes by allowing fact-checkers to corroborate \cite{politico} results using multiple tools and methods \cite{Biden}. Transparent and detailed outputs also contribute to media literacy \cite{restoftheworld}, fostering informed discussions about deceptive AI and enhancing public trust.

The context in which AI detection tools are used significantly affects how their results are interpreted. Confidence scores may be deemed adequate or insufficient depending on the user’s expertise and the specific scenario. As discussed, the individuals often provide the missing context when using the tool and interpret the results within a broader verification process. Therefore, human insight is necessary to spot potential false results. In DRRF, the scores provided by the team’s tools are complemented by analysis of human experts. In a case concerning an audio recording of the Ghanaian President ahead of the last general elections \cite{ghana}, it allowed the team to spot a false positive result produced by their deepfake detection tool. This highlights the necessity of human-assisted detection \cite{NIST}, where analysis is enriched by human expertise. Evaluating tool effectiveness should include metrics such as the time required to obtain results, ease of use for human operators, and comparisons between human-assisted and automated-only detection workflows.

While AI detection tools are generally part of a larger verification ecosystem, certain scenarios demand their use as the primary method for determining authenticity. Cases such as fabricated phone calls or audio recordings, observed by DRRF, illustrate instances where traditional open source investigative techniques, like reverse searches, are ineffective. Personalized and manipulated content of this kind often leaves fact-checkers reliant on AI detection tools as their primary solution. However, recognizing these tools’ limitations in specific disinformation contexts underscores the importance of ensuring they are accessible, reliable, and advanced enough to meet the challenges of such scenarios.

\section{Actionable Steps for Developers and Policy Actors}
Policymakers must support investment in research and development of AI detection, as well as sociotechnical evaluations. They also need to be aware of the balance between providing transparency into technical documentation and the trade-offs this may bring for the security of models and the effectiveness of a tool. Developers, on the other hand, need to collaborate with diverse stakeholders, design tools for real-world challenges, ensure accessibility and fairness, and adhere to ethical standards and transparency.

\subsection{For AI Detection Tools Developers}
The AI developers are encouraged to evaluate AI detection tools against the key considerations stemming from each of the six core pillars of effectiveness. In particular, they are urged to:
\begin{itemize}
    \item Design AI detection solutions that respond to \textbf{real-world challenges} of deepfake detection, including low-quality, compressed, and heavily formatted content, and operate effectively across multiple languages, representations, and cultural contexts, addressing underrepresented demographics in training datasets.
    \item \textbf{Clearly communicate} AI detection results, including expanding on confidence scores, dataset, and tool limitations, to foster credibility and functionality of the tool and ensure that it is accessible to users with varying technical expertise.
    \item Establish internal policies for r\textbf{egular updates, maintenance process, retirement protocols and durability} benchmarks, as well as regularly evaluate the tool performance across different demographics, contexts and against new developments in synthetic media.
    \item Engage with a \textbf{diverse group of stakeholders}, including the AI regulators, standard bodies, and prospective global users, to support development of responsible AI regulations and standards, and ensure that AI detection tools align with principles of trustworthy and human-centric AI.
    \item Conduct \textbf{inclusive and rigorous testing} with diverse teams representative of different stakeholder groups, including external red-teaming, to uncover global regional, contextual blind spots and vulnerabilities
    \item \textbf{Follow WITNESS’ TRIED Benchmark: A Checklist for Truly Innovative and Effective AI Detection} (see the Annex A).
\end{itemize}
\subsection{For International and Domestic Regulatory Bodies}
\begin{itemize}
    \item Incorporate \textbf{sociotechnical considerations} into future regulations, codes of practice and other relevant legislation to ensure that detection tools evaluations reflect real-life applications and user experiences.
    \item Design \textbf{accountability mechanisms} to safeguard fairness and accessibility considerations throughout the whole lifecycle of the detection system while accounting for security and safeguarding against adversarial attacks.
    \item Encourage \textbf{global multistakeholder engagement} to ensure tools reflect diverse needs and contexts. Prioritize collaborative and multistakeholder approach to development of technologies and policies to ensure that the AI detection tools identify and respond to diverse needs and capacities.
\end{itemize}
\subsection{For Standards Bodies}
\begin{itemize}
    \item Set \textbf{minimum standards requirement} for detection tools to be considered truly effective from a social technical perspective based on the considerations outlined in this report.
    \item Establish \textbf{guidelines on transparency, explainability, and fairness} to align detection tools with sociotechnical evaluations and global best practices.
    \item Develop standards for \textbf{regular updates and durability} benchmarks for detection tools to address evolving AI technologies.
\end{itemize}
\subsection{For Governments and Market Leaders}
\begin{itemize}
    \item Push for the \textbf{implementation of detection standards} that ensure that AI detection tools deployed in public-facing contexts meet regulatory, safety and ethical standards.
    \item Provide \textbf{secure and long-term funding mechanisms} that support research, development and long-term maintenance of detection solutions to ensure these can be equitably developed, regularly updated, and successfully adapted to promptly evolving synthetic media technologies.
    \item \textbf{Subsidize access} to detection tools for key and targeted information and human rights actors.
    \item Current and future funs should prioritze investment in \textbf{public interest AI funds} to advance technical solutions that promote a resilient global information ecosystem and facilitate development of local AI resources, expertise and capacities, with a focus on underserved communities and high-risk information environments.
    \item Invest in \textbf{training programs, workshops, and technical assistance} to ensure targeted stakeholders can effectively and responsibly leverage detection tools in real-world contexts.
\end{itemize}
\section{Conclusions}
The threat of deceptive AI continues to grow, posing risks to information integrity and public trust. AI detection tools could play a pivotal role in mitigating the risks of deceptive AI—but only if developed with a sociotechnical evaluation framework. Addressing this challenge requires AI detection tools that are not only technically advanced but that can also adequately support users in diverse, real-world scenarios and complement the existing verification approaches. Ensuring that the AI detection tools are contextually relevant, accessible and fair requires multi-stakeholder commitment and infrastructural support. An holistic approach is needed to address the considerations raised in this report, and both AI developers and policy actors have a key role in advancing detection solutions that foster public trust and strengthen the global information ecosystem. This work has opened up a large scope for future work across many different domains that we encourage civil society, industry, and academia to pursue. WITNESS endeavours to work with partners and collaborators across the recommendations for advocacy and future research and practical implementations.

The \textit{TRIED Benchmark} translates the lessons learnt from the Deepfakes Rapid Response Force interventions into a detailed list of statements encompassing different dimensions of effectiveness discussed throughout the report to ensure that detection tools support and empower communities most impacted by synthetic media threats. In doing so, it effectively bridges the gap between strictly technical evaluations of AI detection systems and their practical applications in high-stake contexts. The Benchmark is designed to guide AI developers, policy actors and other stakeholders in evaluating whether the AI detection tools are effective from a sociotechnical perspective and support the diverse needs of their users while promoting adaptability, transparency, accessibility, fairness, and durability, and strengthening verification workflows.

By adopting the \textit{WITNESS’ TRIED Benchmark} and embracing a comprehensive approach that prioritizes fairness, transparency, and adaptability, stakeholders can ensure detection tools meet the diverse needs of a global audience. This commitment will strengthen the resilience of the information ecosystem, fostering accountability and trust in the age of AI.
\section{Acknowledgments}
We would like to thank all the journalists, fact-checkers and researchers who have referred cases to our Deepfake Rapid Response Force, as well as the over forty experts and members of the Force who continue to provide WITNESS and their partners with pro bono computational models and expertise.

We also wish to thank the following individuals, all of whom provided insightful comments at different stages of the development of the framework: Henry Ajder, (Latent Space Advisory), Rabiu Alhassan (FactSpace West Africa), Mahsa Alimardani (WITNESS), Tracy Aminu (FactSpace West Africa), Ana Bregvadze (GeoFacts), Edward Delp (Purdue College of Engineering), Brandon Epstein (Magnet Forensics), Sam Gregory (WITNESS), Natalie Gyenes (University of Toronto), Rijul Gupta (Deep Media), Gabi Ivens (Human Rights Watch), Sean McGregor (UL Research Institutes), Claire Leibowicz (PAI), Jose Lopez (Intel Labs), Michael Macedo Diniz (Federal Institute of São Paulo), Ryan Ofman (Deep Media), Symeon (Akis) Papadopoulos (CERTH-IT), Maiko Ratiani (Myth Detector), Thomas Roca (Microsoft), Anderson Rocha (University of Campinas), Nikos Sarris (CERTH-ITI), Saniat J. Sohrawardi (Rochester Institute of Technology), Y. Kelly Wu (Rochester Institute of Technology), Raquel Vazquez Llorente (Google), Luisa Verdoliva (University Federico II of Naples). 

We are grateful to the participants of the 9th NYU Computational Disinformation Symposium and the PAI AI and Media Integrity Steering Committee for feedback on this project's development. We also want to thank the reviewers who provided comments but wished to remain anonymous. 
\newpage

\bibliographystyle{plain}
\bibliography{main}

\newpage
\appendix
\section{WITNESS' TRIED Benchmark: A Checklist for Truly Innovative and Effective AI Detection}

\renewcommand{\labelenumii}{\arabic{enumi}.\arabic{enumii}}
\renewcommand{\labelenumiii}{\arabic{enumi}.\arabic{enumii}.\arabic{enumiii}}
\renewcommand{\labelenumiv}{\arabic{enumi}.\arabic{enumii}.\arabic{enumiii}.\arabic{enumiv}}

\large{\textbf{Instructions}}

\normalsize
The checklist below is adapted from the benchmark quality assessment framework outlined in BetterBench by Anka Reuel, Amelia Hardy, Chandler Smith, Max Lamparth, and Mykel Kochenderfer from the Stanford Intelligent Systems Laboratory. Inspired by their methodology, this Benchmark uses a similar checklist format, distributes considerations alongside the stages of the tool’s lifecycle and introduces point scale reflecting different phases of realization of set criteria.  

The checklist is designed to guide AI detection developers in evaluating whether their tools align with best practices for equitable and effective detection throughout the stages of design, development, testing, implementation, and maintenance. Each question in the checklist should be answered with a Yes, No, TODO or N/A, followed by a concise justification (approximately one sentence). Justifications may reference specific page numbers from your paper or include relevant links to additional information. 

Each answer receives a number of points. The points are awarded in the following manner:

\begin{center}
\begin{tabular}{||c c||} 
 \hline
 Answer & Number of points \\ [0.5ex] 
 \hline\hline
 Yes & 3 points \\ 
 \hline
 To do with justification & 2 points \\
 \hline
 To do without justification & 0 points \\
 \hline
 No with justification & 1 point \\
 \hline
 No without justification & 0 points \\ [1ex] 
 \hline
\end{tabular}
\end{center}

Upon completing the checklist, you will have the opportunity to calculate your score manually. At the end, you will be provided with a minimum score to evaluate the overall equitable effectiveness of the tool. These scores serve as a guide to help you assess the quality and rigorousness of your tool’s development process.
\newpage
\LARGE{\textbf{Checklist}}

\Large{\textbf{Benchmark Design}}

\normalsize
\textbf{Design Criteria}
\begin{enumerate}[noitemsep]
    \item The goals, key concepts, primary features, and target audience for the detection tool are clearly outlined.
    \item The design process actively involves input from domain experts with relevant expertise.
    \item Relevant academic research, industry standards, and existing literature are thoroughly reviewed and integrated into the design.
    \item Real-world scenarios and practical use cases are incorporated to guide the tool's development and application.
    \item The mechanisms to ensure the tool’s durability and adaptability to rapid development of synthetic media are defined and prioritized in the design process.
    \item The tool’s funding allows for a responsible and sustainable development and operation of the tool.
\end{enumerate}

\textbf{Checklist}
\begin{enumerate}
    \item The goals, concepts, characteristics, and audience of the detection tool are defined.
    \begin{enumerate}
        \item The intended audience is clearly defined.
                
          $\square \textit{TO DO} \hspace{1cm} \square \textit{YES} \hspace{1cm} \square \textit{NO}\hspace{1cm} \square \textit{N/A} $
        
          Justification:
        \item The use cases on which the tool should be used are clearly described.
                
          $\square \textit{TO DO} \hspace{1cm} \square \textit{YES} \hspace{1cm} \square \textit{NO}\hspace{1cm} \square \textit{N/A} $
        
          Justification:
        \item The goals and the way the detection model works and is designed are explicitly communicated.
                
          $\square \textit{TO DO} \hspace{1cm} \square \textit{YES} \hspace{1cm} \square \textit{NO}\hspace{1cm} \square \textit{N/A} $
        
          Justification:
        \item Differences between this tool and existing detection tools are outlined, including unique capabilities or improvements.
                
          $\square \textit{TO DO} \hspace{1cm} \square \textit{YES} \hspace{1cm} \square \textit{NO}\hspace{1cm} \square \textit{N/A} $
        
          Justification:
    \end{enumerate}
    \item The design process involved consultation with diverse domain experts, including those from global and underserved contexts.
            
          $\square \textit{TO DO} \hspace{1cm} \square \textit{YES} \hspace{1cm} \square \textit{NO}\hspace{1cm} \square \textit{N/A} $
        
          Justification:
    \item Diverse relevant academic research, industry reports, and existing literature were integrated into the tool’s design.
    \begin{enumerate}
        \item The research and literature were diverse and steps were taken to include knowledge produced outside of Europe and the United States.
                
          $\square \textit{TO DO} \hspace{1cm} \square \textit{YES} \hspace{1cm} \square \textit{NO}\hspace{1cm} \square \textit{N/A} $
        
          Justification:
    \end{enumerate}
    \item Real-world scenarios and practical use cases are incorporated to guide the tool's development and application.
    \begin{enumerate}
        \item Real-life cases and use cases were incorporated to reflect practical challenges and expectations.
                
          $\square \textit{TO DO} \hspace{1cm} \square \textit{YES} \hspace{1cm} \square \textit{NO}\hspace{1cm} \square \textit{N/A} $
        
          Justification:
        \item Stakeholders representative of the target audience were consulted in the design phase to clearly outline their key needs and expectations.
                
          $\square \textit{TO DO} \hspace{1cm} \square \textit{YES} \hspace{1cm} \square \textit{NO}\hspace{1cm} \square \textit{N/A} $
        
          Justification:
    \end{enumerate}
    \item The mechanisms to ensure the tool’s durability and adaptability to rapid development of synthetic media are defined.
    \begin{enumerate}
        \item Specific steps are outlined to ensure that the tool is future-proof and guarantee that it will remain relevant in the long term.
                
          $\square \textit{TO DO} \hspace{1cm} \square \textit{YES} \hspace{1cm} \square \textit{NO}\hspace{1cm} \square \textit{N/A} $
        
          Justification:
        \item The tool’s funding allows for a responsible and sustainable development and operation of the tool.
                
                  $\square \textit{TO DO} \hspace{1cm} \square \textit{YES} \hspace{1cm} \square \textit{NO}\hspace{1cm} \square \textit{N/A} $
        
          Justification:
          
    \end{enumerate}
\end{enumerate}

\Large
\begin{table}[ht]
    \centering
    \begin{tabular}{|c|} \hline 
        The maximum number of points is 30. \\  \\Your score is: \\ \hline
    \end{tabular}

\end{table}

\Large{\textbf{Benchmark Development}}

\normalsize
\textbf{Development Criteria}
\begin{enumerate}[noitemsep]
    \item The development stage requires explicit consideration of underrepresented languages, cultural nuances, and geopolitical challenges in both training and testing.
    \item Training data is diverse, representative of global demographics, and collected ethically, with transparent documentation of the sourcing process.
    \item Comprehensive and accessible documentation is provided, integrating relevant context to aid understanding and usability.
    \item The tool's limitations are clearly identified and communicated to ensure realistic expectations of its capabilities.
    \item Mechanisms for explainability are incorporated, ensuring that results and detections are interpretable by both technical and non-technical users.
    \item The development team is diverse, reflecting the perspectives and needs of the tool's intended audience.
    \item Accessibility is prioritized, ensuring the tool is usable by the intended audience regardless of technical expertise or resource constraints.
    \item The tool's capabilities for handling various file types and quality levels are explicitly detailed.
    \item The tool demonstrates consistent performance across diverse contexts and use cases it aims to serve.
    \item Development respects and integrates with existing verification techniques and skill sets to enhance reliability and usability.
    \item Adaptability is a key focus, allowing the tool to evolve in response to new challenges, use cases, and technological advancements.
\end{enumerate}

\textbf{Checklist}
\begin{enumerate}
    \item Training data is diverse, representative of global demographics, and ethically sourced.
    \begin{enumerate}
        \item Data collection process complied with the local data protection regulations.
                        
          $\square \textit{TO DO} \hspace{1cm} \square \textit{YES} \hspace{1cm} \square \textit{NO}\hspace{1cm} \square \textit{N/A} $
        
          Justification:
        \item The training dataset is representative of diverse demographics.
                        
          $\square \textit{TO DO} \hspace{1cm} \square \textit{YES} \hspace{1cm} \square \textit{NO}\hspace{1cm} \square \textit{N/A} $
        
          Justification:
        \item Developers implemented measures to assess and mitigate biases during training.
                        
          $\square \textit{TO DO} \hspace{1cm} \square \textit{YES} \hspace{1cm} \square \textit{NO}\hspace{1cm} \square \textit{N/A} $
        
          Justification:
    \end{enumerate}
    \item The tool's capabilities for handling various file types and quality levels are explicitly detailed.
                    
          $\square \textit{TO DO} \hspace{1cm} \square \textit{YES} \hspace{1cm} \square \textit{NO}\hspace{1cm} \square \textit{N/A} $
        
          Justification:
    \item The training dataset included examples of corrupted synthetic content.
    \begin{enumerate}
        \item The tool was trained on low-resolution content.
                        
          $\square \textit{TO DO} \hspace{1cm} \square \textit{YES} \hspace{1cm} \square \textit{NO}\hspace{1cm} \square \textit{N/A} $
        
          Justification:
        \item The tool was trained on social media compression standards.
                        
          $\square \textit{TO DO} \hspace{1cm} \square \textit{YES} \hspace{1cm} \square \textit{NO}\hspace{1cm} \square \textit{N/A} $
        
          Justification:
        \item The tool was trained on reformatted content, including files resaved in a different format.
                        
          $\square \textit{TO DO} \hspace{1cm} \square \textit{YES} \hspace{1cm} \square \textit{NO}\hspace{1cm} \square \textit{N/A} $
        
          Justification:
        \item If the tool is capable of detecting audio, it was trained on audio content corrupted by noise (including background noise, music and cross-talking)
                        
          $\square \textit{TO DO} \hspace{1cm} \square \textit{YES} \hspace{1cm} \square \textit{NO}\hspace{1cm} \square \textit{N/A} $
        
          Justification:
        \item If the tool is capable of detecting video, it was trained on video content with dynamic movements and noisy background.
                        
          $\square \textit{TO DO} \hspace{1cm} \square \textit{YES} \hspace{1cm} \square \textit{NO}\hspace{1cm} \square \textit{N/A} $
        
          Justification:
    \end{enumerate}
    \item The detection tool was trained to deal with diverse types of content.
    \begin{enumerate}
        \item If the tool is capable of detecting audio, the training dataset included data with a transmission similar to telephones.
                        
          $\square \textit{TO DO} \hspace{1cm} \square \textit{YES} \hspace{1cm} \square \textit{NO}\hspace{1cm} \square \textit{N/A} $
        
          Justification:
        \item If the tool is capable of detecting audio, the training dataset included radio conversations (such as radio broadcasts and walkie-talkie conversations).
                        
          $\square \textit{TO DO} \hspace{1cm} \square \textit{YES} \hspace{1cm} \square \textit{NO}\hspace{1cm} \square \textit{N/A} $
        
          Justification:
        \item If the tool is capable of detecting video, the training dataset included footage featuring dynamic content of various angles and positions, with movements or multiple people.
                        
          $\square \textit{TO DO} \hspace{1cm} \square \textit{YES} \hspace{1cm} \square \textit{NO}\hspace{1cm} \square \textit{N/A} $
        
          Justification:
        \item The tool’s training dataset included different types of files, including file compressions on different social media platforms.
                        
          $\square \textit{TO DO} \hspace{1cm} \square \textit{YES} \hspace{1cm} \square \textit{NO}\hspace{1cm} \square \textit{N/A} $
        
          Justification:
    \end{enumerate}
    \item The tool is developed to scale across different levels of content complexity, size, and volume.
    \begin{enumerate}
        \item A system is in place, and adaptable to adjust the detection model to continually update the training dataset to include new examples of AI-generated or manipulated content.
                        
          $\square \textit{TO DO} \hspace{1cm} \square \textit{YES} \hspace{1cm} \square \textit{NO}\hspace{1cm} \square \textit{N/A} $
        
          Justification:
    \end{enumerate}
    \item The tool provides users with clear information on how detection results should be interpreted.
    \begin{enumerate}
        \item The tool does not use binary labels (such as ‘real’ or ‘fake’) when communicating the results but instead describes manipulation using accessible description and language of degrees.
                        
          $\square \textit{TO DO} \hspace{1cm} \square \textit{YES} \hspace{1cm} \square \textit{NO}\hspace{1cm} \square \textit{N/A} $
        
          Justification:
        \item The tool includes validation processes that go beyond confidence scores, offering meaningful contextual insights or probability distributions.
                        
          $\square \textit{TO DO} \hspace{1cm} \square \textit{YES} \hspace{1cm} \square \textit{NO}\hspace{1cm} \square \textit{N/A} $
        
          Justification:
        \item The additional information outlines the tool’s capabilities, including what content it can analyze, top level information on the data it was trained on, and the training objectives.
                        
          $\square \textit{TO DO} \hspace{1cm} \square \textit{YES} \hspace{1cm} \square \textit{NO}\hspace{1cm} \square \textit{N/A} $
        
          Justification:
        \item The additional information outlines the tool's limitations, including the language constraints and the challenges that may impact its performance, such as the quality of analyzed content.
                        
          $\square \textit{TO DO} \hspace{1cm} \square \textit{YES} \hspace{1cm} \square \textit{NO}\hspace{1cm} \square \textit{N/A} $
        
          Justification:
        \item The tool points to the specific elements indicating that the content is synthetic, such as the exact moment when the manipulation occurred, specific regions identified as synthetic, or metadata-based insights supporting its conclusion.
                        
          $\square \textit{TO DO} \hspace{1cm} \square \textit{YES} \hspace{1cm} \square \textit{NO}\hspace{1cm} \square \textit{N/A} $
        
          Justification:
    \end{enumerate}
    \item The tool acknowledges other existing verification techniques and incorporates information about them into its workflow.
    \begin{enumerate}
        \item The information is provided with respect to other verification techniques and skill sets that could be integrated.
                        
          $\square \textit{TO DO} \hspace{1cm} \square \textit{YES} \hspace{1cm} \square \textit{NO}\hspace{1cm} \square \textit{N/A} $
        
          Justification:
        \item When the detection tool cannot provide a conclusive result, the accompanying information still provides valuable insights to the user and may be used along with other verification methods.
                        
          $\square \textit{TO DO} \hspace{1cm} \square \textit{YES} \hspace{1cm} \square \textit{NO}\hspace{1cm} \square \textit{N/A} $
        
          Justification:
    \end{enumerate}
    \item The tool is adaptable with processes in place to evolve in response to new challenges, use cases, and technological advancements.
                    
          $\square \textit{TO DO} \hspace{1cm} \square \textit{YES} \hspace{1cm} \square \textit{NO}\hspace{1cm} \square \textit{N/A} $
        
          Justification:
\end{enumerate}
\large
\Large
\begin{table}[ht]
    \centering
    \begin{tabular}{|c|} \hline 
        The maximum number of points is 57. \\  \\Your score is: \\ \hline
    \end{tabular}
\end{table}

\newpage
\Large{\textbf{Benchmark Testing}}

\normalsize
\textbf{Testing Criteria}
\begin{enumerate}[noitemsep]
    \item Proactive measurements are being taken to test the tool’s blind spots.
    \item The tool undergoes regular and systematic testing to maintain reliability and effectiveness.
    \item Diverse stakeholders and domain experts, including independent external groups, actively participate in the testing process to provide varied perspectives and expertise.
    \item The tool is rigorously tested on challenging edge cases, including adversarial content, false claims of AI-generated media, and heavily manipulated content.
    \item The tool demonstrates resilience against evasion techniques, maintaining its accuracy and reliability even under deliberate attempts to bypass detection.
    \item The tool is evaluated from a human-detection perspective.
\end{enumerate}

\textbf{Checklist}
\begin{enumerate}
    \item The tool was tested in an exhaustive and inclusive manner.
    \begin{enumerate}
        \item The tool is proactively tested to identify blind spots and potential failures.
                            
          $\square \textit{TO DO} \hspace{1cm} \square \textit{YES} \hspace{1cm} \square \textit{NO}\hspace{1cm} \square \textit{N/A} $
        
          Justification:
        \item The tool’s robustness was tested against adversarial attacks and evasion techniques.
                            
          $\square \textit{TO DO} \hspace{1cm} \square \textit{YES} \hspace{1cm} \square \textit{NO}\hspace{1cm} \square \textit{N/A} $
        
          Justification:
        \item External stakeholders were involved in the red-teaming of the tool.
                            
          $\square \textit{TO DO} \hspace{1cm} \square \textit{YES} \hspace{1cm} \square \textit{NO}\hspace{1cm} \square \textit{N/A} $
        
          Justification:
        \item The tool was tested by a global group of stakeholders from different regions to identify vulnerabilities and challenges stemming from application of the tool in specific cultural contexts.
                            
          $\square \textit{TO DO} \hspace{1cm} \square \textit{YES} \hspace{1cm} \square \textit{NO}\hspace{1cm} \square \textit{N/A} $
        
          Justification:
        \item Testing is conducted to evaluate the tool's performance on real-world examples of deceptive and manipulated content.
                            
          $\square \textit{TO DO} \hspace{1cm} \square \textit{YES} \hspace{1cm} \square \textit{NO}\hspace{1cm} \square \textit{N/A} $
        
          Justification:
        \item Results and performance metrics are documented, and top level information is transparent.
                            
          $\square \textit{TO DO} \hspace{1cm} \square \textit{YES} \hspace{1cm} \square \textit{NO}\hspace{1cm} \square \textit{N/A} $
        
          Justification:
    \end{enumerate}
    \item The tool is evaluated from a human-assisted detection perspective.
    \begin{enumerate}
        \item Such testing included assessing how much time it took to receive the result and how difficult the process was from the human perspective.
                            
          $\square \textit{TO DO} \hspace{1cm} \square \textit{YES} \hspace{1cm} \square \textit{NO}\hspace{1cm} \square \textit{N/A} $
        
          Justification:
        \item Such testing included a comparison between the performance of human-assisted detection and detection by an individual not using a detection tool.
                            
          $\square \textit{TO DO} \hspace{1cm} \square \textit{YES} \hspace{1cm} \square \textit{NO}\hspace{1cm} \square \textit{N/A} $
        
          Justification:
    \end{enumerate}
\end{enumerate}

\Large
\begin{table}[ht]
    \centering
    \begin{tabular}{|c|} \hline 
        The maximum number of points is 24. \\  \\Your score is: \\ \hline
    \end{tabular}
\end{table}

\Large{\textbf{Benchmark Implementation}}

\normalsize
\textbf{Implementation Criteria}
\begin{enumerate}[noitemsep]
    \item The tool is designed and implemented to uphold human rights, ensuring it does not infringe on privacy, freedom of expression, or other fundamental rights.
    \item Accessibility is prioritized, ensuring the tool is usable by its intended audience, regardless of technical expertise or resource availability.
    \item Documentation is clear and comprehensive, and top level version is easily accessible, providing users with the necessary guidance to operate the tool effectively and understand its limitations.
\end{enumerate}

\textbf{Checklist}
\begin{enumerate}
    \item Ensure the tool does not inadvertently violate human rights, such as privacy or freedom of expression.
    \begin{enumerate}
        \item The tool explicitly avoids outputs or recommendations that could harm individuals or communities, aligning with ethical AI principles.
                                    
          $\square \textit{TO DO} \hspace{1cm} \square \textit{YES} \hspace{1cm} \square \textit{NO}\hspace{1cm} \square \textit{N/A} $
        
          Justification:
        \item The tool ensures secure handling of sensitive data and complies with relevant data protection regulations.
                                    
          $\square \textit{TO DO} \hspace{1cm} \square \textit{YES} \hspace{1cm} \square \textit{NO}\hspace{1cm} \square \textit{N/A} $
        
          Justification:
        \item The tool ensures that it does not discriminate in its outputs and that its accuracy does not vary depending on the demographic of the individuals featured in the content.
                                    
          $\square \textit{TO DO} \hspace{1cm} \square \textit{YES} \hspace{1cm} \square \textit{NO}\hspace{1cm} \square \textit{N/A} $
        
          Justification:
    \end{enumerate}
    \item The tool is accessible to a diverse group of targeted users.
    \begin{enumerate}
        \item The tool includes educational resources or tutorials to help users understand its functionality and limitations.
                                    
          $\square \textit{TO DO} \hspace{1cm} \square \textit{YES} \hspace{1cm} \square \textit{NO}\hspace{1cm} \square \textit{N/A} $
        
          Justification:
        \item The tool does not require advanced technical knowledge or skills from the user to operate the tool.
                                    
          $\square \textit{TO DO} \hspace{1cm} \square \textit{YES} \hspace{1cm} \square \textit{NO}\hspace{1cm} \square \textit{N/A} $
        
          Justification:
        \item The tool operates efficiently without requiring significant computational power or strong network connectivity.
                                    
          $\square \textit{TO DO} \hspace{1cm} \square \textit{YES} \hspace{1cm} \square \textit{NO}\hspace{1cm} \square \textit{N/A} $
        
          Justification:
        \item The tool has the option to function offline.
                                    
          $\square \textit{TO DO} \hspace{1cm} \square \textit{YES} \hspace{1cm} \square \textit{NO}\hspace{1cm} \square \textit{N/A} $
        
          Justification:
        \item Multilingual support is available for target audiences, and limitations in other language support are clearly communicated.
                                    
          $\square \textit{TO DO} \hspace{1cm} \square \textit{YES} \hspace{1cm} \square \textit{NO}\hspace{1cm} \square \textit{N/A} $
        
          Justification:
        \item The tool is affordable to the targeted group of users.
                                    
          $\square \textit{TO DO} \hspace{1cm} \square \textit{YES} \hspace{1cm} \square \textit{NO}\hspace{1cm} \square \textit{N/A} $
        
          Justification:
        \item The tool provides clear, coherent summaries of its results tailored to a general audience.
                                    
          $\square \textit{TO DO} \hspace{1cm} \square \textit{YES} \hspace{1cm} \square \textit{NO}\hspace{1cm} \square \textit{N/A} $
        
          Justification:
    \end{enumerate}
    \item The tool’s performance metrics are accurate and not exaggerated.
    \begin{enumerate}
        \item The tool communicates errors or uncertainties clearly to users, avoiding overconfidence in ambiguous cases.
                                    
          $\square \textit{TO DO} \hspace{1cm} \square \textit{YES} \hspace{1cm} \square \textit{NO}\hspace{1cm} \square \textit{N/A} $
        
          Justification:
    \end{enumerate}
\end{enumerate}

\large
\Large
\begin{table}[ht]
    \centering
    \begin{tabular}{|c|} \hline 
        The maximum number of points is 33. \\  \\Your score is: \\ \hline
    \end{tabular}
\end{table}
\newpage
\Large{\textbf{Benchmark Maintenance}}

\normalsize
\textbf{Maintenance Criteria}

\begin{enumerate}[noitemsep]
    \item The tool undergoes regular performance evaluations to ensure its detection accuracy remains reliable across evolving datasets and content types.
    \item Resources are actively being distributed for updates and maintenance processes.
    \item Updates are routinely implemented to address emerging challenges, such as adversarial attacks, advancements in deepfake technologies, and new forms of synthetic content.
    \item User feedback is actively collected, documented, and incorporated into updates, with a transparent and accessible communication channel available for issue reporting and suggestions.
    \item Clear policies are established regarding the support duration for older versions of the tool following updates or new releases.
    \item Periodic audits are conducted to identify and mitigate any biases introduced during updates or revealed by newer datasets.
    \item Developers continuously monitor advancements in AI and related technologies, ensuring the tool remains aligned with the latest innovations and best practices.
\end{enumerate}

\textbf{Checklist}
\begin{enumerate}
    \item The tool undergoes regular updates, and changes are documented transparently.
    \begin{enumerate}
        \item The tool’s accuracy is evaluated against previously verified cases.
                                            
          $\square \textit{TO DO} \hspace{1cm} \square \textit{YES} \hspace{1cm} \square \textit{NO}\hspace{1cm} \square \textit{N/A} $
        
          Justification:
        \item Clear policies are in place for how regularly updates will be conducted.
                                            
          $\square \textit{TO DO} \hspace{1cm} \square \textit{YES} \hspace{1cm} \square \textit{NO}\hspace{1cm} \square \textit{N/A} $
        
          Justification:
        \item The tool is periodically reviewed for its applicability and usability across diverse cultural and linguistic contexts.
                                            
          $\square \textit{TO DO} \hspace{1cm} \square \textit{YES} \hspace{1cm} \square \textit{NO}\hspace{1cm} \square \textit{N/A} $
        
          Justification:
        \item Regular audits are conducted to detect and address bias introduced in updates or uncovered in newer datasets.
                                            
          $\square \textit{TO DO} \hspace{1cm} \square \textit{YES} \hspace{1cm} \square \textit{NO}\hspace{1cm} \square \textit{N/A} $
        
          Justification:
        \item The tool communicates to the user when it was last updated.
                                            
          $\square \textit{TO DO} \hspace{1cm} \square \textit{YES} \hspace{1cm} \square \textit{NO}\hspace{1cm} \square \textit{N/A} $
        
          Justification:
        \item Documentation is kept up to date with every tool revision, including new use cases, limitations, and changes in features.
                                            
          $\square \textit{TO DO} \hspace{1cm} \square \textit{YES} \hspace{1cm} \square \textit{NO}\hspace{1cm} \square \textit{N/A} $
        
          Justification:
    \end{enumerate}
    \item An accessible channel of communication is provided.
    \begin{enumerate}
        \item Feedback channels for users are actively maintained, and updates incorporate user feedback.
                                            
          $\square \textit{TO DO} \hspace{1cm} \square \textit{YES} \hspace{1cm} \square \textit{NO}\hspace{1cm} \square \textit{N/A} $
        
          Justification:
        \item A contact person or team for support and inquiries is clearly listed.
                                            
          $\square \textit{TO DO} \hspace{1cm} \square \textit{YES} \hspace{1cm} \square \textit{NO}\hspace{1cm} \square \textit{N/A} $
        
          Justification:
        \item An active community of users, developers, and domain experts is cultivated to provide ongoing support and collaboration.
                                            
          $\square \textit{TO DO} \hspace{1cm} \square \textit{YES} \hspace{1cm} \square \textit{NO}\hspace{1cm} \square \textit{N/A} $
        
          Justification:
    \end{enumerate}
    \item The tool has End-of-Life procedures in place.
    \begin{enumerate}
        \item The tool will be retired if it has not been updated for a specified time following clear policies.
                                            
          $\square \textit{TO DO} \hspace{1cm} \square \textit{YES} \hspace{1cm} \square \textit{NO}\hspace{1cm} \square \textit{N/A} $
        
          Justification:
        \item If the tool is to be discontinued, clear communication is provided to users, along with data migration or alternative solutions.
                                            
          $\square \textit{TO DO} \hspace{1cm} \square \textit{YES} \hspace{1cm} \square \textit{NO}\hspace{1cm} \square \textit{N/A} $
        
          Justification:
    \end{enumerate}
\end{enumerate}

\large
\Large
\begin{table}[ht]
    \centering
    \begin{tabular}{|c|} \hline 
        The maximum number of points is 33. \\  \\Your score is: \\ \hline
    \end{tabular}
\end{table}

\begin{table}[ht]
    \centering
    \begin{tabular}{|c|} \hline 
        \Large\textbf{The Total Maximum Number of Points is 177}. \\  \\ \Large A score above 143 points indicates that your tool is \\ \Large\textbf{truly effective}. \\  \\ \Large A score between 107 and 142 points indicates that your tool is \\ \Large\textbf{moderately effective}. \\ \\ \Large A score between 72 and 106 points indicates that your tool is \\ \Large\textbf{somewhat effective}.  \\  \\ \Large A score below 71 points indicates that your tool is \\ \Large\textbf{not effective}.  \\ \hline
    \end{tabular}
\end{table}

\end{document}